\documentclass[12pt, letterpaper]{article}

\usepackage{amsmath,amssymb,amsfonts}
\usepackage{algorithmic}
\usepackage{algorithm}
\usepackage{array}
\usepackage[caption=false,font=normalsize,labelfont=sf,textfont=sf]{subfig}
\usepackage{textcomp}
\usepackage{stfloats}
\usepackage{url}
\usepackage{verbatim}
\usepackage{float}
\usepackage{graphicx}
\usepackage{multirow}
\usepackage{dirtytalk}
\usepackage{cite}
\usepackage{color}

\usepackage{url}

\makeatletter

\begin{document}

\title{SPICE-PIDE: A Methodology for Design and Optimization of Integrated Circuits}
\author{Jehan Taraporewalla\textsuperscript{1,2,$\dagger$}, Arun KP\textsuperscript{1,$\dagger$}, Sugata Ghosh\textsuperscript{1}, Abhishek Agarwal\textsuperscript{1},\\ 
Bijaydoot Basak\textsuperscript{1}, and Dipankar Saha\textsuperscript{1,3}}
\date{\textsuperscript{1}Department of Electronics and Telecommunication Engineering, Indian Institute of Engineering Science and Technology Shibpur, Howrah-711103, India\\
\textsuperscript{2}Department of Electrical Engineering, Indian Institute of Technology Delhi, New Delhi-110016, India\\ 
\textsuperscript{3}Department of Electronics and Communication Engineering, Narula Institute of Technology, Agarpara, Kolkata-700109, India.\\
\textsuperscript{$\dagger$}J.T and A. KP contributed equally to work\\
e-mail: dipsah\_etc@yahoo.co.in and pc.jehan@gmail.com}
\maketitle              
\newpage
\begin{abstract}
In application-specific designs, owing to the trade-off between power consumption and speed, optimization of various circuit parameters has become a challenging task. Several of the performance metrics, viz., energy efficiency, gain, performance, and noise immunity, are interrelated and difficult to tune. Such efforts may result in a great deal of manual iterations which in turn increase the computational overhead. Thus, it is important to develop a methodology that not only explores large design space but also reduces the computational time. In this work, we investigate the viability of using a SPICE and Python IDE (PIDE) interface to optimize integrated circuits. The SPICE simulations are carried out using 22 nm technology node with a nominal supply voltage of 0.8 V. The SPICE-PIDE optimizer, as delineated in this work, is able to provide the best solution sets considering various performance metrics and design complexities for 5 transistor level converters.
\end{abstract}


\section{Introduction}
Optimization of integrated circuits is a difficult task for the designer owing to the trade-offs among various parameters viz. supply voltage, gain, linearity, power consumption, output swing, noise, etc. \cite{Moradi,Razavi,Zele,Settaluri}. In the manual approach, while designing any integrated circuit, the designer has to size each component individually, and that takes a long time \cite{Huang,Saha,Mukherjee}. Therefore, we require an accurate method which is compatible with industry standard EDA tools such as SPICE engine \cite{Zele,Wolfe}.\\

\begin{figure}[!h]
    \centering
\includegraphics[width=0.535\linewidth]{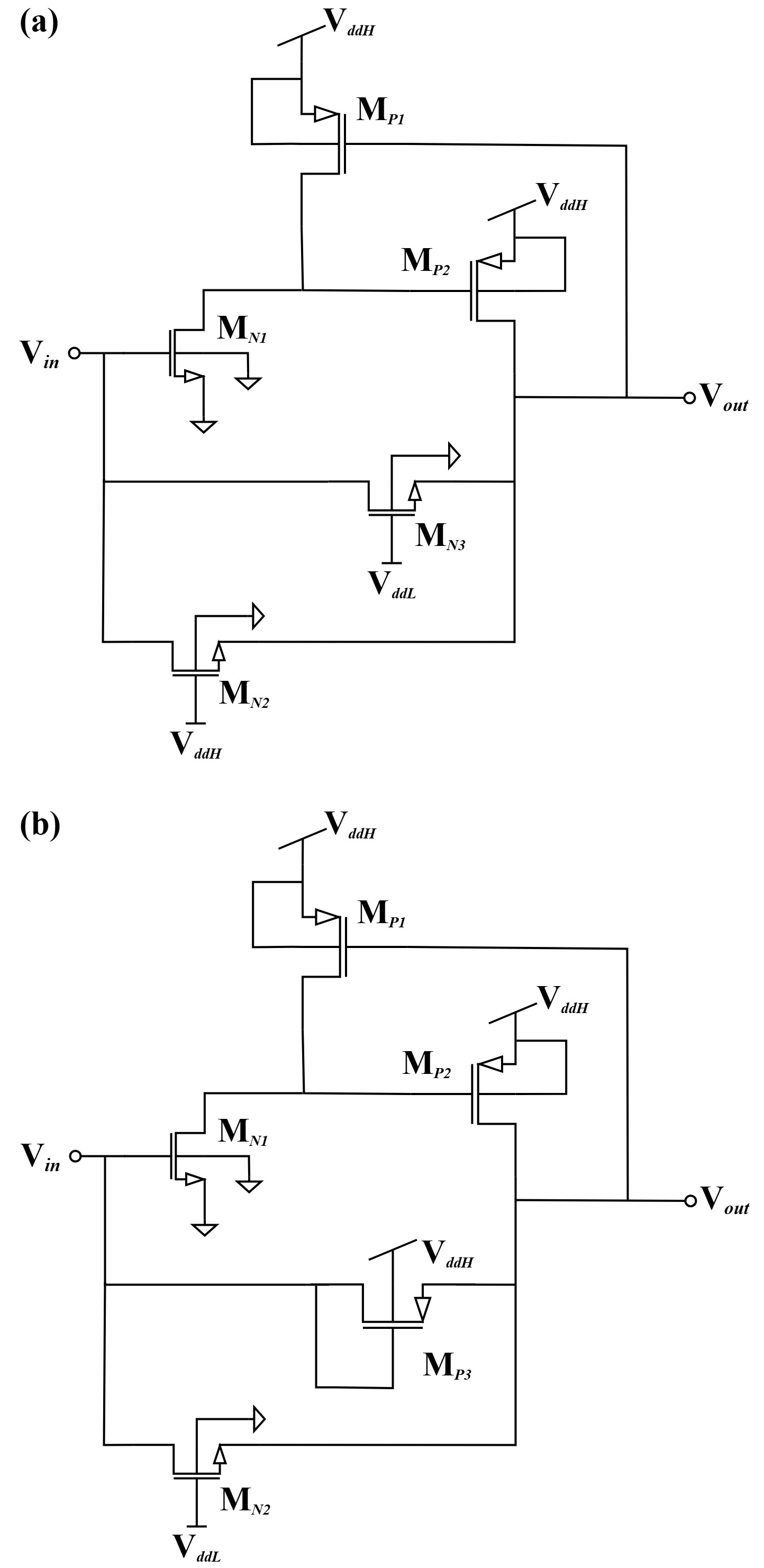}
    \caption{Architecture of 5 transistor level converters (a) NNPT and (b) PNPT with the transistor sizing of 40nm/40nm for M\textsubscript{P2} and load of 5 fF. The V\textsubscript{ddH} and V\textsubscript{ddL} values are 0.8 V and 0.6 V respectively \cite{NNPT}.}
    \label{Fig_1_NNPT_PNPT}
\end{figure}

The goal of this work is to examine the viability of Python IDE for design optimization via data analysis, parameter selection, and prioritization of efficient solutions \cite{Ghosh}. In this work, we consider the 5 transistor level converter designs reported in \cite{NNPT} and employ SPICE-PIDE \cite{Ghosh} in order to obtain optimal solutions with ultra-low power consumption. These 5 transistor ultra-low power level converters (as shown in Fig. \ref{Fig_1_NNPT_PNPT}) can work effectively in the near-threshold region with energy efficiency at par with other sub-threshold level converters \cite{LC_indicon,Vishwani,Hosseini_TCS,Hui_Shao}. We carry out a detailed study that involves the exploration of design space and considers 1843968 cases in order to find a set of best solutions during the optimization of 5 transistor level converters.

\section{Methodology}
In order to predict the behaviour of the 5 transistor level converters in state of the art technology node, we employ the metal gate, high-K, and strained-Si High Performance 22 nm Predictive Technology Model (PTM) version: 4.0 level: 54 \cite{PTM}. All simulations are done using SPICE with 22 nm technology node \cite{LTspice}.
\begin{figure}[!h]
    \begin{center}
    \includegraphics[width=0.625\linewidth]{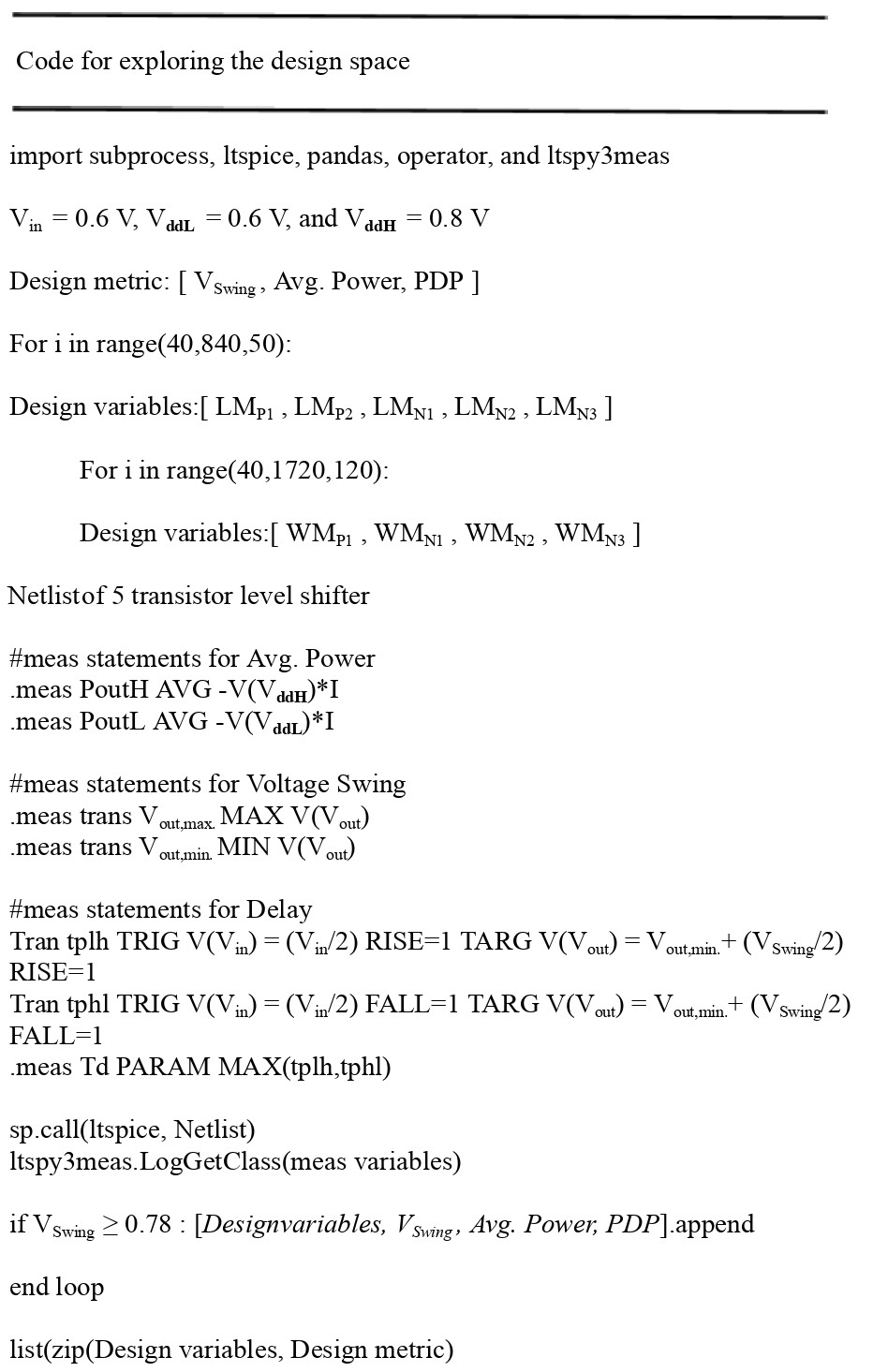}
    \caption{Pseudo code for optimization of NNPT LC using SPICE-PIDE.}
    \label{Fig_2_Code}
        \end{center}
\end{figure}
\begin{figure}[!h]
    \begin{center}
    \includegraphics[width=0.625\linewidth]{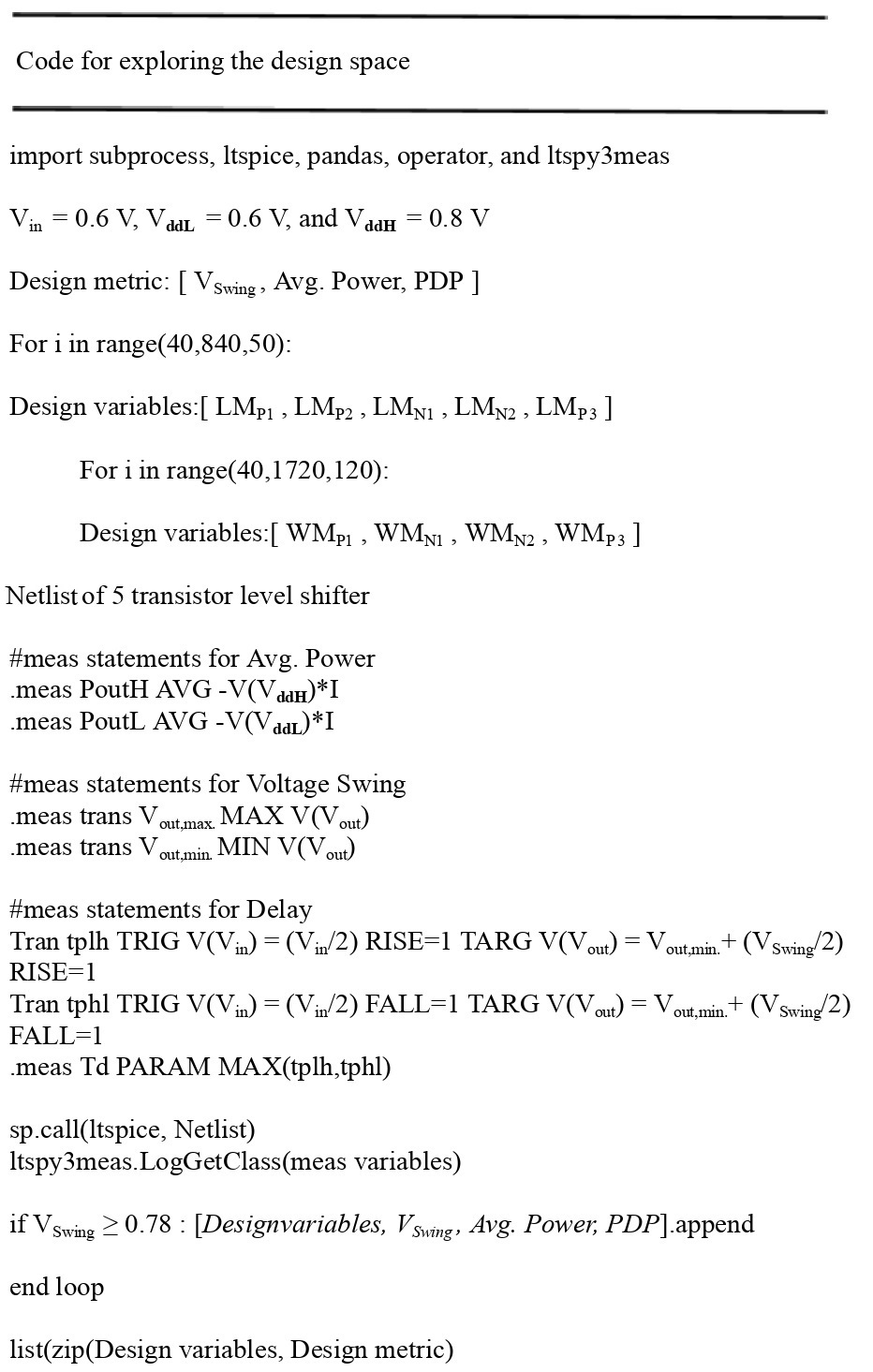}
    \caption{Pseudo code for optimization of PNPT LC using SPICE-PIDE.}
    \label{Fig_3_Code}
        \end{center}
\end{figure}

The default NMOS and PMOS transistors as per our recipe for carrying out the simulations have length and width values: L\textsubscript{NMOS}= L\textsubscript{PMOS}= 40 nm, W\textsubscript{NMOS}= 200 nm, and W\textsubscript{PMOS}= 500 nm. Moreover, the nominal supply voltage reported in High Performance 22 nm Predictive Technology Model is 0.8 V with the reported threshold voltages of 0.503 V $\&$ 0.460 V, for NMOS and PMOS transistors \cite{PTM,PTM_ASU}. A 10 ns rise time / fall time is taken for the input pulse and the contribution of both supplies (V\textsubscript{ddL} and V\textsubscript{ddH}) is considered while calculating the average power consumption. Besides, a 5 fF load capacitance (C\textsubscript{L}) is used for all the simulations, carried out at a frequency of 10 MHz.
\begin{figure}[!h]
\centering
    \includegraphics[width=0.5\linewidth]{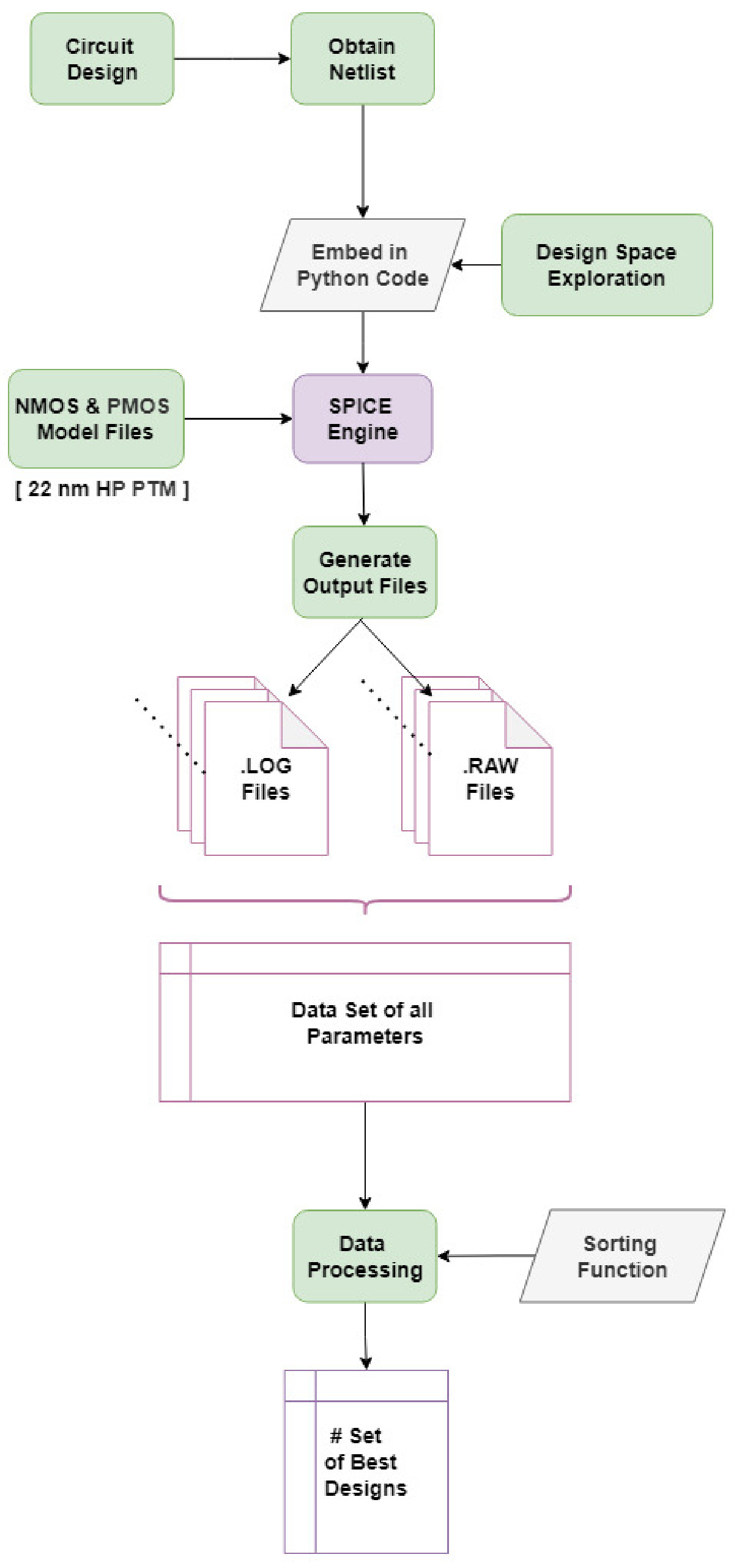}
    \caption{Flow diagram illustrating the design optimization of 5 transistor level converters.}
    \label{Fig_4_Flowchart}
\end{figure}

The SPICE-PIDE optimizer as discussed in \cite{Ghosh}, heuristically explores the design space and lists all possible solutions for any circuit design. The codes for exploring the design space of NNPT and PNPT level converters are delineated in Fig. \ref{Fig_2_Code} and \ref{Fig_3_Code}, respectively. It utilizes data processing code to find out the best solution set considering various key performance parameters. The Fig. \ref{Fig_4_Flowchart} shows the flow diagram depicting integration between SPICE and PIDE utilized for design optimization of level converters. It emphasises on the various steps taken by SPICE-PIDE optimizer to obtain the set of best solutions.

\section{Results}
Here, the SPICE-PIDE optimizer is employed to explore umpteen possible solutions for ultra low power signal conversion utilizing the SPICE engine \cite{LTspice} in conjunction with Spyder (a well known Python IDE) \cite{Spyder}. The combination of parameters used to investigate the possible solutions are as follows:
\begin{itemize}
    \item Input voltage, V\textsubscript{in} = 0.6 V
    \item  Supply voltage, High, V\textsubscript{ddH} = 0.8 V
    \item  Supply voltage, Low, V\textsubscript{ddL} = 0.6 V
    \item W\textsubscript{P2} = 40 nm (is not varied)
    \item Transistor length values (L\textsubscript{i}) are varied across 16 values from 40 nm to 840 nm
    \item All other transistor width values (W\textsubscript{i}) are varied across 14 values from 40 nm to 1720 nm
     \end{itemize}    
   
To evaluate and rank the various solutions, several performance metrics e.g. output voltage swing (V\textsubscript{Swing}), average power dissipation (P\textsubscript{Avg.}, considering V\textsubscript{ddH} and V\textsubscript{ddL}), and Power Delay Product (PDP) are used.\\ 

It is to be noted that the design space for NNPT level converter considers the combination of two gate voltages viz. V\textsubscript{G\textunderscore N2} = V\textsubscript{ddH} \& V\textsubscript{G\textunderscore N3} = V\textsubscript{ddL} and V\textsubscript{G\textunderscore N2} = V\textsubscript{ddL} \& V\textsubscript{G\textunderscore N3} = V\textsubscript{ddH} respectively during evaluation. The tables (Table I \& Table II) illustrate the best solution sets considering the performance parameters as discussed earlier. It is evident from the results, as listed in Table I, that the average power consumption is significantly reduced (26.5866 nW), compared to 32.5704 nW as reported in \cite{NNPT} (while maintaining complete output voltage swing). Considering interfacing with other digital blocks via long interconnects, such a robust and energy efficient level converter design can be very useful.

\begin{table}[!h]\tiny
\centering
\label{NNPT_Avg._Power_Table}
\caption{Solution set ranked based on P\textsubscript{Avg.} while optimizing LC NNPT}
 \begin{tabular}{p{3.75cm}p{2.35cm}p{1cm}p{1cm}p{1cm}p{1cm}} 
 \hline
 &&&&&\\
\footnotesize\textbf{Solution Set} & \footnotesize \textbf{Gate Voltage Combination} &  \footnotesize \textbf{P\textsubscript{Avg.} (nW)} &  \footnotesize \textbf{T\textsubscript{dmax} (ns)} & \footnotesize \textbf{V\textsubscript{out\textunderscore low} (nV)}& \footnotesize \textbf{V\textsubscript{out\textunderscore high} (mV)}\\
 &&&&&\\
 \hline \hline  
 &&&&&\\
\scriptsize {L\textsubscript{i}=90nm W\textsubscript{P1}=160nm W\textsubscript{N1}=40nm W\textsubscript{N2}=640nm W\textsubscript{N3}=160nm}& \scriptsize{V\textsubscript{G\textunderscore N2} = V\textsubscript{ddH} V\textsubscript{G\textunderscore N3} = V\textsubscript{ddL}}& \scriptsize 26.5866 & \scriptsize 0.624 & \scriptsize 408.923 & \scriptsize 799.984\\
 &&&&&\\
\scriptsize {L\textsubscript{i}=90nm W\textsubscript{P1}=160nm W\textsubscript{N1}=40nm W\textsubscript{N2}=160nm W\textsubscript{N3}=640nm}&\scriptsize V\textsubscript{G\textunderscore N2} = V\textsubscript{ddL} V\textsubscript{G\textunderscore N3} = V\textsubscript{ddH}&\scriptsize 26.5866&\scriptsize 0.624 &\scriptsize 408.924&\scriptsize 799.984\\
&&&&&\\
\scriptsize {L\textsubscript{i}=140nm W\textsubscript{P1}=40nm W\textsubscript{N1}=40nm W\textsubscript{N2}=40nm W\textsubscript{N3}=760nm}& \scriptsize V\textsubscript{G\textunderscore N2} = V\textsubscript{ddL} V\textsubscript{G\textunderscore N3} = V\textsubscript{ddH} &\scriptsize 26.5892&\scriptsize 0.763&\scriptsize 1988.65&\scriptsize 799.981\\
&&&&&\\
\scriptsize {L\textsubscript{i}=90nm W\textsubscript{P1}=40nm W\textsubscript{N1}=40nm W\textsubscript{N2}=160nm W\textsubscript{N3}=400nm}& \scriptsize V\textsubscript{G\textunderscore N2} = V\textsubscript{ddL} V\textsubscript{G\textunderscore N3} = V\textsubscript{ddH} &\scriptsize 26.5893&\scriptsize 0.697&\scriptsize 1026.84&\scriptsize 799.99\\
&&&&&\\
\scriptsize {L\textsubscript{i}=90nm W\textsubscript{P1}=40nm W\textsubscript{N1}=40nm W\textsubscript{N2}=400nm W\textsubscript{N3}=160nm}& \scriptsize V\textsubscript{G\textunderscore N2} = V\textsubscript{ddH} V\textsubscript{G\textunderscore N3} = V\textsubscript{ddL} &\scriptsize 26.5893&\scriptsize 0.697&\scriptsize 1026.84&\scriptsize 799.99\\
&&&&&\\
\scriptsize {L\textsubscript{i}=140nm W\textsubscript{P1}=40nm W\textsubscript{N1}=40nm W\textsubscript{N2}=760nm W\textsubscript{N3}=40nm}& \scriptsize V\textsubscript{G\textunderscore N2} = V\textsubscript{ddH} V\textsubscript{G\textunderscore N3} = V\textsubscript{ddL} &\scriptsize 26.6465&\scriptsize 0.763&\scriptsize 1979.61&\scriptsize 799.981\\
&&&&&\\
\scriptsize {L\textsubscript{i}=90nm W\textsubscript{P1}=40nm W\textsubscript{N1}=40nm W\textsubscript{N2}=640nm W\textsubscript{N3}=40nm}& \scriptsize V\textsubscript{G\textunderscore N2} = V\textsubscript{ddH} V\textsubscript{G\textunderscore N3} = V\textsubscript{ddL} &\scriptsize 26.7466&\scriptsize 0.647 &\scriptsize 503.803&\scriptsize 799.925\\ 
&&&&&\\
\scriptsize {L\textsubscript{i}=90nm W\textsubscript{P1}=160nm W\textsubscript{N1}=40nm W\textsubscript{N2}=40nm W\textsubscript{N3}=520nm}& \scriptsize V\textsubscript{G\textunderscore N2} = V\textsubscript{ddL} V\textsubscript{G\textunderscore N3} = V\textsubscript{ddH} &\scriptsize 26.7486&\scriptsize 0.641&\scriptsize 1281.01&\scriptsize 799.942\\
\hline 
\end{tabular}
\end{table}

Howbeit, if we prioritize the performance and consider the metric PDP, then the results presented in Table II exhibit a little improvement (i.e. 8.528 $\times 10^{-18}$ J as compared to 9.412 $\times 10^{-18}$ J as reported in \cite{NNPT}). Moreover, the aforementioned solutions do not significantly increase the silicon area overhead.\\  

\begin{table}[h]\tiny
\centering
\label{NNPT_PDP_Table}
\caption{Solution set ranked based on PDP while optimizing LC NNPT}

 \begin{tabular}{p{3.75cm}p{2.35cm}p{1cm}p{1cm}p{1cm}p{1cm}} 
 \hline
 &&&&&\\

\footnotesize \textbf{Solution Set} & \footnotesize \textbf{Gate Voltage Combination} &  \footnotesize \textbf{PDP (aJ)} &  \footnotesize \textbf{T\textsubscript{dmax} (ns)}  & \footnotesize \textbf{V\textsubscript{out\textunderscore low} (nV)}& \footnotesize \textbf{V\textsubscript{out\textunderscore high} (mV)}\\
 &&&&&\\
 \hline \hline  
&&&&&\\
\scriptsize{L\textsubscript{i}=40nm W\textsubscript{P1}=40nm W\textsubscript{N1}=160nm W\textsubscript{N2}=1000nm W\textsubscript{N3}=40nm}& \scriptsize{V\textsubscript{G\textunderscore N2} = V\textsubscript{ddH} V\textsubscript{G\textunderscore N3} = V\textsubscript{ddL}} &\scriptsize 8.528&\scriptsize 0.254  &\scriptsize 177.783 &\scriptsize 799.967\\
&&&&&\\
\scriptsize {L\textsubscript{i}=40nm W\textsubscript{P1}=40nm W\textsubscript{N1}=160nm W\textsubscript{N2}=40nm W\textsubscript{N3}=1000nm}&\scriptsize V\textsubscript{G\textunderscore N2} = V\textsubscript{ddL} V\textsubscript{G\textunderscore N3} = V\textsubscript{ddH}&\scriptsize 8.594&\scriptsize 0.254&\scriptsize 177.779&\scriptsize 799.967\\
&&&&&\\

\scriptsize {L\textsubscript{i}=40nm W\textsubscript{P1}=40nm W\textsubscript{N1}=160nm W\textsubscript{N2}=1480nm W\textsubscript{N3}=40nm}& \scriptsize V\textsubscript{G\textunderscore N2} = V\textsubscript{ddH} V\textsubscript{G\textunderscore N3} = V\textsubscript{ddL} &\scriptsize 8.692&\scriptsize 0.249&\scriptsize 176.226&\scriptsize 799.949\\
&&&&&\\

\scriptsize {L\textsubscript{i}=40nm W\textsubscript{P1}=40nm W\textsubscript{N1}=160nm W\textsubscript{N2}=40nm W\textsubscript{N3}=1480nm}& \scriptsize V\textsubscript{G\textunderscore N2} = V\textsubscript{ddL} V\textsubscript{G\textunderscore N3} = V\textsubscript{ddH} &\scriptsize 8.692&\scriptsize 0.249&\scriptsize 176.226&\scriptsize 799.949\\
&&&&&\\

\scriptsize {L\textsubscript{i}=40nm W\textsubscript{P1}=160nm W\textsubscript{N1}=400nm W\textsubscript{N2}=1120nm W\textsubscript{N3}=40nm}& \scriptsize V\textsubscript{G\textunderscore N2} = V\textsubscript{ddH} V\textsubscript{G\textunderscore N3} = V\textsubscript{ddL} &\scriptsize 8.725&\scriptsize 0.254&\scriptsize 188.121&\scriptsize 799.962\\
&&&&&\\
\scriptsize {L\textsubscript{i}=40nm W\textsubscript{P1}=40nm W\textsubscript{N1}=160nm W\textsubscript{N2}=1480nm W\textsubscript{N3}=160nm}& \scriptsize V\textsubscript{G\textunderscore N2} = V\textsubscript{ddH} V\textsubscript{G\textunderscore N3} = V\textsubscript{ddL} &\scriptsize 8.832&\scriptsize 0.251&\scriptsize 175.812&\scriptsize 799.951\\
\hline 
\end{tabular}
\end{table}
\newpage
As listed in Table III $\&$ Table IV, the solutions exhibit improved average power consumption (36.5228 nW) and PDP (19.98$\times 10^{-18}$ J) which are $\sim 16.6\%$ and $\sim 5.5\%$ improvement respectively compared to that of the PNPT level converter described in \cite{NNPT}. Moreover, the PNPT level converter presented here demonstrates a complete restoration of the output voltage, though with a slightly larger silicon area.\\  

\begin{table}[!h]\tiny
\centering
\caption{Solution set ranked based on P\textsubscript{Avg.} while optimizing LC PNPT}
 \begin{tabular}{p{5cm}p{1cm}p{1cm}p{1cm}p{1cm}p{1cm}} 
 \hline
&&&&&\\
\footnotesize \textbf{Solution Set}  &  \footnotesize \textbf{P\textsubscript{Avg.} (nW)} &  \footnotesize \textbf{T\textsubscript{dmax} (ns)} & \footnotesize \textbf{PDP (aJ)} & \footnotesize \textbf{V\textsubscript{out\textunderscore low} (nV)}& \footnotesize \textbf{V\textsubscript{out\textunderscore high} (mV)}\\
&&&&&\\
 \hline \hline  
&&&&&\\
\scriptsize {L\textsubscript{i}=90nm W\textsubscript{P1}=40nm W\textsubscript{N1}=40nm W\textsubscript{N2}=760nm W\textsubscript{P3}=40nm}&\scriptsize 36.5228&\scriptsize 1.086 &\scriptsize 39.663&\scriptsize 346.467 &\scriptsize 799.991\\
&&&&&\\
\scriptsize {L\textsubscript{i}=140nm W\textsubscript{P1}=40nm W\textsubscript{N1}=40nm W\textsubscript{N2}=1000nm W\textsubscript{P3}=160nm}&\scriptsize 36.5776&\scriptsize 1.859&\scriptsize 67.997&\scriptsize 525.352&\scriptsize 799.956\\
&&&&&\\
\scriptsize {L\textsubscript{i}=190nm W\textsubscript{P1}=40nm W\textsubscript{N1}=40nm W\textsubscript{N2}=1000nm W\textsubscript{P3}=40nm} &\scriptsize 37.2161&\scriptsize 2.555&\scriptsize 95.087&\scriptsize 1605.65&\scriptsize 800.047\\
&&&&&\\
\scriptsize {L\textsubscript{i}=240nm W\textsubscript{P1}=40nm W\textsubscript{N1}=40nm W\textsubscript{N2}=1480nm W\textsubscript{P3}=40nm} &\scriptsize 37.2264&\scriptsize 3.291&\scriptsize 122.512&\scriptsize 2583.5&\scriptsize 799.956\\
&&&&&\\
\scriptsize {L\textsubscript{i}=190nm W\textsubscript{P1}=40nm W\textsubscript{N1}=40nm W\textsubscript{N2}=1240nm W\textsubscript{P3}=40nm} &\scriptsize 37.3237&\scriptsize 2.529&\scriptsize 94.391&\scriptsize 2147.98&\scriptsize 799.988\\
&&&&&\\ 
\scriptsize {L\textsubscript{i}=240nm W\textsubscript{P1}=40nm W\textsubscript{N1}=40nm W\textsubscript{N2}=1000nm W\textsubscript{P3}=40nm} &\scriptsize 37.5053&\scriptsize 3.272&\scriptsize 122.714&\scriptsize 1735.81&\scriptsize 800.003\\
&&&&&\\
\scriptsize {L\textsubscript{i}=40nm W\textsubscript{P1}=40nm W\textsubscript{N1}=760nm W\textsubscript{N2}=640nm W\textsubscript{P3}=40nm} &\scriptsize 37.76&\scriptsize 1.093&\scriptsize 47.271&\scriptsize 812.5&\scriptsize 799.992\\
\hline 
\end{tabular}
\end{table}
\begin{table}[!h]\tiny
\centering
\caption{Solution set ranked based on PDP while optimizing LC PNPT}
 \begin{tabular}{p{5cm}p{1cm}p{1cm}p{1cm}p{1cm}p{1cm}}
 \hline
&&&&&\\
\footnotesize \textbf{Solution Set}  &  \footnotesize \textbf{P\textsubscript{Avg.} (nW)} &  \footnotesize \textbf{T\textsubscript{dmax} (ns)} & \footnotesize \textbf{PDP (aJ)} & \footnotesize \textbf{V\textsubscript{out\textunderscore low} (nV)}& \footnotesize \textbf{V\textsubscript{out\textunderscore high} (mV)}\\
&&&&&\\
 \hline \hline  
 &&&&&\\ 
\scriptsize {L\textsubscript{i}=40nm W\textsubscript{P1}=40nm W\textsubscript{N1}=760nm W\textsubscript{N2}=880nm W\textsubscript{P3}=40nm} &\scriptsize 47.4606&\scriptsize 0.421 &\scriptsize 19.980&\scriptsize 919.093 &\scriptsize 799.993\\
&&&&&\\ 
\scriptsize {L\textsubscript{i}=40nm W\textsubscript{P1}=40nm W\textsubscript{N1}=640nm W\textsubscript{N2}=880nm W\textsubscript{P3}=40nm}&\scriptsize 47.4487&\scriptsize 0.432&\scriptsize 20.497&\scriptsize 80.336&\scriptsize 799.991\\
&&&&&\\ 
\scriptsize {L\textsubscript{i}=40nm W\textsubscript{P1}=40nm W\textsubscript{N1}=520nm W\textsubscript{N2}=640nm W\textsubscript{P3}=40nm} &\scriptsize 46.1263&\scriptsize 0.445&\scriptsize 20.526&\scriptsize 80.155&\scriptsize 799.976\\
&&&&&\\ 
\scriptsize {L\textsubscript{i}=40nm W\textsubscript{P1}=40nm W\textsubscript{N1}=520nm W\textsubscript{N2}=880nm W\textsubscript{P3}=40nm} &\scriptsize 46.2314&\scriptsize 0.445&\scriptsize 20.572&\scriptsize 76.848&\scriptsize 799.989\\
&&&&&\\ 
\scriptsize {L\textsubscript{i}=40nm W\textsubscript{P1}=40nm W\textsubscript{N1}=880nm W\textsubscript{N2}=880nm W\textsubscript{P3}=40nm} &\scriptsize 49.8017&\scriptsize 0.414&\scriptsize 20.617&\scriptsize 83.068&\scriptsize 799.989\\
&&&&&\\ 
\scriptsize {L\textsubscript{i}=40nm W\textsubscript{P1}=40nm W\textsubscript{N1}=760nm W\textsubscript{N2}=1000nm W\textsubscript{P3}=40nm}&\scriptsize 49.1144&\scriptsize 0.422&\scriptsize 20.726&\scriptsize 411.976&\scriptsize 799.989\\
&&&&&\\ 
\scriptsize {L\textsubscript{i}=40nm W\textsubscript{P1}=40nm W\textsubscript{N1}=760nm W\textsubscript{N2}=640nm W\textsubscript{P3}=40nm} &\scriptsize 49.565&\scriptsize 0.419&\scriptsize 20.767&\scriptsize 82.857&\scriptsize 799.988\\
&&&&&\\ 
\scriptsize {L\textsubscript{i}=40nm W\textsubscript{P1}=40nm W\textsubscript{N1}=640nm W\textsubscript{N2}=1000nm W\textsubscript{P3}=40nm} &\scriptsize 48.3256&\scriptsize 0.433&\scriptsize 20.924&\scriptsize 82.184&\scriptsize 799.988\\
\hline 
\end{tabular}
\end{table}

\newpage
\section{Conclusion}
In this paper, we employ SPICE-PIDE to optimize integrated circuits that operate with dual supply voltages.
We find that the SPICE-PIDE optimizer can provide the solution sets considering metrics like avg. power, PDP, etc. for any circuit without increasing the computational load significantly. Besides, the most efficient solutions, as shown in Table I $\&$ Table II, can be utilized for the purpose of designing ultra-low power systems with improved performance.

\setcounter{secnumdepth}{0}
\section{Acknowledgements}
Arun KP thanks the department of Electronics and Telecommunication Engineering, Indian Institute of Engineering Science and Technology, Shibpur, India, for all the facilities and resources.

\end{document}